\documentclass{aastex}
\usepackage[dvips]{epsfig}
\usepackage{array}

\newcommand{\ts}{\thinspace}

\shorttitle{Hard Tails in Sco~X-1}
\shortauthors{D'Amico et al.}

\begin{document}
\title{\bf HEXTE DETECTIONS OF HARD X--RAY TAILS IN SCO X--1}

\author{Flavio D'Amico{\altaffilmark{1,2}}, William A. Heindl{\altaffilmark{1}}, \\
        Richard E. Rothschild{\altaffilmark{1}}, and Duane E. Gruber{\altaffilmark{1}}}
\email{fdamico@ucsd.edu, wheindl@ucsd.edu, rrothschild@ucsd.edu, dgruber@ucsd.edu}
\altaffiltext{1}{University of California, San Diego -- UCSD \\
                 9500 Gilman Drive, La Jolla, CA 92093-0424}
\altaffiltext{2}{Instituto Nacional de Pesquisas Espaciais -- INPE \\
                 Av. dos Astronautas 1758, 12227-010 S.J. dos Campos - SP, Brazil}

\begin{abstract}
We report the detection of a non--thermal hard X--ray component from Sco X--1 based upon
the analysis of 20--220{\ts}keV spectra obtained with the HEXTE experiment on-board the {\it RXTE}
satellite. We find that the addition of a power--law component to a thermal bremsstrahlung
model is required to achieve a good fit in 5 of 16 observations analyzed. Using PCA data we were
able to track the movement of the source along the {\sf Z} diagram, 
and we found that the presence of the hard
X--ray tail is not confined to a specific {\sf Z} position.
However, we do observe an indication that the power law index  hardens
with increasing ${\dot{M}}$, as indicated from the position on the {\sf Z} diagram. 
We find that the derived non--thermal luminosities are $\sim${\ts}10{\%} of that derived for 
the brightest of the atoll sources.

\end{abstract}


\section{Introduction}\label{intro}

Observations of hard X--ray emission from atoll sources (e.g. Barret, 
McClintock, {\&} Grindlay 1996; Barret et al. 2000) have
revealed the unexpected result that power law spectra extending
above 30{\ts}keV may not be an exclusive signature of a black hole.
{\it Rossi X--ray Timing Explorer (RXTE)} observations of hard X--ray emission from 
SAX{\ts}J1808.4$-$3658 (Heindl {\&} Smith 1998) demonstrated that hard X--ray flux is also present
in other types of low--mass X--ray binaries (LMXBs).
These observations refined the role of hard X--ray emission in distinguishing between
neutron star and black hole binaries. 

Scorpius X--1 is a high luminosity LMXB {\sf Z} source,
where the primary is a neutron star with a low magnetic field. 
The presence of a non--thermal component in Sco X--1 has
been suggested (e.g., Peterson {\&} Jacobson 1966; Riegler, Bolt, {\&}
Serlemitsos 1970; Agrawal et al. 1971; Haymes et al. 1972; Duldig et al. 1983)
and the absence of such a component has also
been reported (e.g., Lewin, Clark, {\&} Smith 1967; Buselli et al. 1968; Jain et al. 1973;
Johnson et al. 1980). In fact, high sensitivity searches have
failed to detect such a component, placing strong upper limits on the non-thermal
flux (e.g., Greenhill et al. 1979; Rothschild et al. 1980; Soong {\&} Rothschild 1983).
More recently, however, Strickman {\&} Barret (2000)
have reported the presence of a hard X--ray tail in Sco X--1 using {\it CGRO}/OSSE.

We used {\it RXTE} observations from the public
database to search for hard X--ray tails in Sco X--1 spectra. Sco X--1 has been
observed with {\it RXTE} several times, and our data sample (1997 April  to 1999 July) contains
185,672{\ts}s of Proportional Counter Array (PCA) on--source time and
more than 100{\ts}ks of on--source live-time in the High Energy X-ray Timing
Experiment (HEXTE). We describe data selection and analysis in
\S{\ts}{\ref{data}}, results in \S{\ts}{\ref{results}}, discuss the 
detection of a non-thermal component in 
\S{\ts}{\ref{disc}}, and present our conclusions in \S{\ts}{\ref{conc}}.

\section{Data Selection and Analysis}\label{data}

In this work we used data from HEXTE to search for hard X--ray tails in Sco X--1
in the $\sim${\ts}20--220{\ts}keV interval, and data from the PCA to determine 
the {\sf Z} position of Sco X--1 in the color--color diagram. 
The PCA (Jahoda et al. 1996) is a set of 5 identical Xenon proportional counters covering
the energy range 2--60{\ts}keV with a total area of
$\sim${\ts}7000{\ts}cm$^2$. The HEXTE (Rothschild et al. 1998) consists of two clusters of 4
NaI(Tl)/CsI(Na) phoswich scintillation counters totaling
$\sim${\ts}1600{\ts}cm$^2$ of effective area in the 15--250{\ts}keV
energy range . Each cluster 
rocks between source and two background fields to measure the instrumental background
in near real time. The PCA background is estimated from particle rates and
the known diffuse X--ray flux. The PCA and HEXTE share a common 1$^{\circ}$ full
width at half maximum field of view. Pertinent to this work, the HEXTE continuum
sensitivity at 100{\ts}keV, for $\sim${\ts}5{\ts}kseconds of live--time, 
is $\sim${\ts}13{\ts}mCrab (3$\sigma$, scaled to in-flight backgrounds from the 
predictions of Gruber et al. 1996).  

We selected those observations from the {\it RXTE} public database (as of 2000 April)
in which $\gtrsim${\ts}5000{\ts}s
of HEXTE total on--source time is available, in order to achieve good sensitivity at
high energies. Since some observations did not point directly at Sco X--1 (in order
to reduce the very high counting rate), we have also discarded observations for which
{\it RXTE} pointing offsets were larger than 0.01$^{\circ}$. These criteria resulted in
selection of the 16 observations displayed in Table {\ref{tab1}}. 
The resulting accumulated on--source HEXTE live--time is 104,238{\ts}s.

\subsection{Data Analysis}

We performed our analysis using HEXTE {\tt Archive Histogram
Mode} data, which are accumulated every 16{\ts}s, have spectral
information in 64 compressed channels, and
are present for all observations, independent of selected Science Mode. 

The data analysis was performed using software from {\tt FTOOLS 5.0}.
For each observation, we divided the data into subsets according to
position on the {\sf Z} diagram (see below and Table {\ref{tab1}}).  We then
accumulated individual source and background count histograms for HEXTE
clusters A and B and added them to form a single HEXTE source and
background histogram for each subset. Response matrices for HEXTE A and B
were generated and added (with weights of 0.57 and 0.43 respectively,
according to their effective areas). 

In order to be confident that any high energy excesses were real, we
tailored our data selection to optimize background subtraction.  We
verified that any remaining systematic errors as a result of our
procedure were small and did not affect the results. We followed the
standard practices extensively studied for HEXTE background
subtraction by us and other workers (Rothschild et al. 1998; MacDonald
1999).  First, we selected only data taken more than 15{\ts}minutes
after satellite passage through the South Atlantic Anomaly (SAA).
This allows any short--lived radioactive species produced in the
detector material to decay and ensures that the systematic error of
our linearly interpolated background estimate is
$\lesssim${\ts}0.02{\ts}{\%} of the background. We verified this
process by comparing -- for each cluster -- the spectra from the two
separate background fields which are sampled alternately during
observations. In each case, subtracting these spectra showed no excess
emission, giving us confidence in the on--source background
estimate. This also verified that there were no confusing sources in the
background fields. We note, too, that none of the Sco X--1 spectra
show evidence for the prominent instrumental background lines at
$\sim${\ts}30, 70, 160 and 190{\ts}keV (Rothschild et al. 1998) which
we would expect to be apparent if there were over or under--subtracted
backgrounds. Even though there was no obvious sign of residual
background features, we allowed the normalization of the background
spectra to vary in our fitting procedure. As expected, the fitted
adjustment to the background level was always
$\lesssim${\ts}1{\ts}{\%}. Even with this added degree of freedom, the
significance of the detected high energy tails remained essentially
unchanged. This also verified that live-times were
calculated accurately, as systematic errors in the exposure would lead
to incorrect background subtraction. Finally, at 100{\ts}keV (see
Figure {\ref{fig1}}), the Sco X--1 hard flux is of the order of
0.02{\ts}counts/s/keV, while the HEXTE background was typically
0.23{\ts}counts/s/keV. Thus, our measurements are at level of
$\sim${\ts}10{\ts}{\%} of the background. MacDonald (1999) showed that
our background subtraction method, even for very long integrations, is
free of systematic errors at a level of $\sim${\ts}0.05{\ts}{\%} of
the background.  A more detailed discussion of HEXTE background
systematics is given in Gruber et al. (1996).

We used the PCA data in order to constrain the position of Sco X--1 in a
color--color diagram. High resolution ($\sim$ 4{\ts}ms) lightcurves 
provided power spectra to characterize any quasi-periodic oscillations and/or 
continuum noise. Low resolution (16{\ts}s) lightcurves were used
to generate color--color diagrams. We defined soft color as the ratio of 
the counts in the (5--7)/(3--5){\ts}keV energy bands and hard color as the ratio
(7--20)/(5--7){\ts}keV. For observations 5/6, 12--15, 17/18 and 20/21 
(see Table {\ref{tab1}}) the source was moving between the Normal and Flaring 
Branches. For those 5 dual--branch observations, we divided each into Normal 
and Flaring Branch subsets. This then resulted in 21 subsets.

\subsection{Spectral Fitting}

We used {\tt XSPEC 11.0} to fit the 20--220{\ts}keV HEXTE spectra 
with a combination of a {\it thermal} component
plus a {\it hard} component. To fit the
thermal component we  tested the following models:
thermal Comptonization (e.g. Lamb {\&} Sanford 1979),
blackbody, and thermal bremsstrahlung. The thermal bremsstrahlung
model returned, in all of our subsets, the smallest ${\chi}^2$,
and it was utilized to represent the lower energy HEXTE flux.

In the 5 subsets in which we detected the presence of a hard X--ray tail
(see next section), a simple power law model, as well as ones with a spectral
break, were evaluated. We found that the simple power law model
described the hard component well, and it was used to represent the higher energy HEXTE flux.
In all 5 subsets with significant non--thermal flux, a high temperature thermal component
did not provide an acceptable fit.

\section{Results}\label{results}

We developed two criteria to determine the presence of a hard X--ray tail in a
particular spectrum: 1) a signal to noise ratio (SNR) $\geq${\ts}5 in the
75--220{\ts}keV energy range, and 2) an F--Test null significance for the addition 
of the hard component at a level of 10$^{-7}$ or less. 
We claim that
we observed a hard X--ray tail {\bf only} when both these criteria were
fulfilled. We thus have 5 strong detections of non--thermal flux
in Sco X--1 spectra, with the hard X--ray tail extending at least to
220{\ts}keV. The spectral parameters derived for those subsets are shown in
Table {\ref{tab2}}. 
Similarly we defined a strong non--detection when 
1) the SNR in the 75--220{\ts}keV is $<${\ts}1 and 2) the F--Test is at a level 
of $\gtrsim${\ts}10$^{-5}$. We thus have 4 strong non--detections.
We note that several subsets formally have very significant F--Test values, but 
SNR values $<${\ts}5. 
These typically result from deviations in the 30--50{\ts}keV range,  
possibly due to imperfect modeling of the thermal component.
We cannot conclude that these subsets have truly non--thermal emission,
although this possibility is open. In Figure
{\ref{fig1}} we show spectra both from one of our detections and one of
our non--detections.

The derived photon indices ($\Gamma$) for the 5 subsets where the tail was
observed spanned a wide range: from $-$0.7 to 2.4, which differs from the
OSSE results which have an average of 2.5 (Strickman {\&} Barret 2000).
It is suggestive that the two hardest photon indices were obtained when
the source was on the flaring branch, and the softest were on the horizontal
branch, showing a possible correlation between the hardness of the spectrum
and the mass accretion rate ${\dot{M}}$, which increases in the sequence
HB{\ts}$\rightarrow${\ts}NB{\ts}$\rightarrow${\ts}FB (e.g., van der Klis 1996).
We also note that both times the source was on the HB a hard X--ray tail was 
detected. 

The non--thermal 20--200{\ts}keV flux from the 5 detections varied by a factor 
of $\sim${\ts}2.5 from 
0.63---1.56{\ts}$\times${\ts}10$^{-9}${\ts}ergs{\ts}cm$^{-2}${\ts}s$^{-1}$ 
(see Table{\ts}{\ref{tab2}}). Upper limits to the 20--200{\ts}keV flux from the 4 
strong non--detections were estimated using a photon index of 1, the mean from the 
5 detections. The one non--detection subset with equivalent sensitivity to the 
detection subsets has a 2$\sigma$ upper limit to the 20--200{\ts}keV flux of 
1.0{\ts}$\times${\ts}10$^{-10}${\ts}ergs{\ts}cm$^{-2}${\ts}s$^{-1}$. This implies 
at least an order of magnitude in overall variability in the non-thermal component 
in Sco X--1.

In order to estimate the 2--20{\ts}keV luminosity, $L^{total}_{2-20}$,
we used a complex multicomponent model to heuristically fit the PCA spectra.
We estimate the uncertainty in the derived flux due to uncertainties in the
proper spectral shape to be $\sim${\ts}2{\ts}{\%}.
Using 2.8{\ts}kpc as the distance to Sco X--1 (Bradshaw, Fomalont, \& Geldzahler 1999),
$L^{total}_{2-20}$ ranges over 2.1--3.2{\ts}$\times${\ts}10$^{38}$ ergs
s$^{-1}$ for the 5 detection subsets. Similarly, the range of 
the 20--200 keV luminosities
of the non--thermal component measurements is
$L^{nt}_{20-200}${\ts}$=${\ts}5.9--15.0{\ts}$\times${\ts}10$^{35}${\ts}ergs{\ts}s$^{-1}$.
Thus the non--thermal component represents about 1{\ts}{\%} of the total
flux from Sco X--1.

No correlation between the {\bf presence} of the hard tail and
the position of the source in the color--color diagram is indicated,
i.e., we have observed the hard X--ray tail in all three branches.
In addition the HEXTE thermal component is not correlated with the presence of the hard
X--ray tail, i.e., the presence of the hard tail is indicated
by neither the temperature nor flux of the HEXTE thermal component. 

\section{Discussion}\label{disc}

From the lack of correlation between the presence of the hard tail and
the position of the source in the color--color diagram
source it would appear
that the hard X--ray tail emission region is not associated
with that part of the accretion disk which is believed to be
responsible for the QPO behavior (van der Klis et al. 1996).
These results also do not support a  correlation between the presence of the hard tail 
and  clustering in the color--color diagram, such as mentioned by Strickman 
{\&} Barret (2000). We also note that this behavior is different from GX{\ts}17{\ts}$+${\ts}2, 
in which a hard tail is correlated with the position in the {\sf Z} (Di Salvo et al. 2000).
Frontera et al. (1998) have also reported the detection of a hard tail in another {\sf Z}
source (Cyg X--2).

Nevertheless we must point out that the observed photon indices may be related 
to the position in the {\sf Z} diagram. The hardest observed photon indices are
in the FB and the softest ones in the HB. From this point of view the production of
hard X--ray photons may be correlated with the accretion rate at the inner part of the disk.
Motivated by the observation that  Sco X--1 is a variable radio source 
(e.g., Fender {\&} Hendry 2000), we considered whether or not the hard X--ray emission 
could be explained in terms of synchrotron emission. While the measured photon 
indices are consistent with this idea for the HB and NB, the ones measured
when the source was in the FB (consistent with 0) are not, since it would require 
an inverted distribution of relativistic electrons. Similarly, the HB and NB spectra 
could result from Comptonization, but the FB power law indices are again too hard. 

A difference between atoll sources and Sco X--1 is also seen in the soft X-ray
luminosity. The atoll source hard tails appear when the 2--20{\ts}keV luminosity
is relatively low, $<${\ts}10$^{37}${\ts}ergs{\ts}s$^{-1}$; whereas the soft
flux in Sco X--1 is essentially at the Eddington limit 
($\sim${\ts}10$^{38}${\ts}ergs s$^{-1}$). It is also interesting to note
that some models used to explain the hard X--ray emission of atoll LMXBs 
(e.g. Mitsuda et al. 1989; Hanawa, 1990) are applicable in a low luminosity state 
up to $\sim${\ts}20{\ts}keV. This, however, is not the case of our 
observations. On the other hand, Sco X--1 may be similar to some atoll sources in 
the luminosity of the power law component. The Sco X--1 power law luminosity in 
the 20--200{\ts}keV range is comparable to the weaker atoll source hard X-ray 
luminosities reported by Barret, McClintock, {\&} Grindlay (1996).

Distinguishing neutron stars from black holes remains a key goal in
high energy astrophysics. Recently it has
been argued (Barret et al. 2000) that the luminosities in the
1--20{\ts}keV (defined as L$_x$) and 20--200{\ts}keV (defined as L$_{hx}$) 
bands could be used to distinguish between 
black hole and low state neutron star systems displaying hard X--ray emission.
In their argument, the critical  luminosity, L$_{crit}$ =
1.5{\ts}$\times${\ts}10$^{37}${\ts}ergs{\ts}s$^{-1}$, is the
dividing line between neutron star and black hole binaries with
the neutron star systems emitting L$_{hx}$ $\sim$ L$_x$ when
L$_x$ $\lesssim$ L$_{crit}$. Since Sco X--1 is emitting Eddington--level soft fluxes,
we cannot test that idea.
On the other hand, our observed L$_{hx}$ does not contradict the idea 
pointed out by Barret et al. (2000)  that only black hole binaries can have both 
L$_x$ and L$_{hx}$ above L$_{crit}$, since our 
measured L$_{hx}$ is below the L$_{crit}$ value (see Table{\ts}{\ref{tab1}}).

\section{Conclusion}\label{conc}

We have observed significant detections of a non--thermal
component in Sco X--1 on 5 out of 21 occasions with HEXTE on-board {\it RXTE}.
We also have found that the presence of the hard X--ray
tail is not associated with a particular position in the color--color
diagram, but that the photon indices, instead, may be correlated with
the position on the {\sf Z} diagram, albeit with a very limited number of examples.
We inferred that the non--thermal power law
flux of the source varied by over an order of magnitude
comparing our detections with the non--detections,
and by a factor of $\sim${\ts}3 in the 5 cases where the presence of the
hard X--ray tail was strongly indicated. 
Our observed 20--200 keV luminosities are in agreement with  only
black hole binaries  emitting hard X--rays with
luminosity $\gtrsim$1.5{\ts}$\times$10$^{37}${\ts}ergs{\ts}s$^{-1}$.

\acknowledgements

This research has made use of data obtained through the
HEASARC, provided by the NASA/GSFC. FD gratefully acknowledges
FAPESP/Brazil for financial support under grant 99/02352-2. FD
acknowledges Stefan Dieters, Wayne Coburn and John Tomsick for helpful discussions.
We also thanks an anonymous referee for suggestions. 
This research was supported by NASA contract NAS5--30720.

\clearpage

\begin{table}
{\footnotesize{
\begin{center}
\caption{Selected {\it RXTE} data observations of Sco X--1 \label{tab1}}
\begin{tabular}{c c c c c c c c c}
\tableline
\tableline
  OBSID           &  MJD  & Nu.& Z Pos.&  Live--time{\tablenotemark{a}}  & SNR{\tablenotemark{b}} & F$_{test}{\tablenotemark{c}}$ & Flux{\tablenotemark{d}} & Hard Tail (Y/N) \\
\tableline
20053-01-01-00    & 50556 &  1 &  HB   &  6341  & 10.6  & 4.4E$-$15  & 8.57$^{+1.52}_{-1.38}$ & Y  \\
20053-01-01-02    & 50558 &  2 &  NB   &  5434  &  4.2  & 1.7E$-$7   & 4.05$^{+0.57}_{-0.87}$ &    \\
20053-01-01-03    & 50559 &  3 &  NB   &  5896  &  4.8  & 3.9E$-$9   & 5.20$^{+0.75}_{-0.74}$ &    \\
20053-01-01-05    & 50561 &  4 &  NB   &  1908  &  0.8  & 6.2E$-$5   & 6.29$^{+1.77}_{-1.77}$ & N  \\
20053-01-01-05    & 50561 &  5 &  FB   &  4593  &  2.4  & 7.7E$-$4   & 2.47$^{+1.94}_{-1.51}$ &    \\
20053-01-01-06    & 50562 &  6 &  NB   &  8558  &  7.3  & 3.5E$-$13  & 8.63$^{+0.91}_{-0.90}$ & Y  \\
10061-01-03-00    & 50815 &  7 &  NB   &  5484  &  4.2  & 2.7E$-$9   & 5.48$^{+0.69}_{-0.74}$ &    \\
20053-01-02-00    & 50816 &  8 &  FB   &  8145  & 18.6  & 5.8E$-$12  & 13.6$^{+2.0}_{-2.0}$   & Y  \\
20053-01-02-03    & 50819 &  9 &  NB   &  5686  &  3.5  & 2.7E$-$9   & 5.91$^{+0.96}_{-0.95}$ &    \\
30036-01-01-000   & 50820 & 10 &  FB   &  6858  &  5.8  & 3.8E$-$7   & 7.73$^{+1.62}_{-1.65}$ & Y  \\
20053-01-02-04    & 50820 & 11 &  FB   &  4193  &  4.9  & 3.3E$-$4   & 3.44$^{+1.23}_{-1.00}$ &    \\
30036-01-02-000   & 50821 & 12 &  NB   &  4370  &  2.2  & 3.6E$-$4   & 4.50$^{+0.87}_{-0.90}$ &    \\
30036-01-02-000   & 50821 & 13 &  FB   &  2527  &  1.8  & 3.6E$-$4   & 2.83$^{+1.05}_{-1.06}$ &    \\
30406-01-02-00    & 50872 & 14 &  NB   &  2326  &  2.5  & 1.0E$-$6   & 4.03$^{+1.07}_{-1.13}$ &    \\
30406-01-02-00    & 50872 & 15 &  FB   &  3125  & $<$ 0 & 0.08       & $<$ 2.20               & N  \\
30035-01-01-00    & 50963 & 16 &  HB   &  6253  &  6.2  & 1.2E$-$14  & 3.63$^{+0.92}_{-0.88}$ & Y  \\
30035-01-03-00    & 50965 & 17 &  NB   &  3750  &  1.3  & 2.4E$-$8   & 3.20$^{+0.86}_{-0.89}$ &    \\
30035-01-03-00    & 50965 & 18 &  FB   &  2318  & $<$ 0 & 0.03       & $<$ 2.67               & N  \\
30035-01-04-00    & 50966 & 19 &  FB   &  6230  & $<$ 0 & 0.38       & $<$ 1.00               & N  \\
40706-01-01-000   & 51433 & 20 &  NB   &  8076  &  2.8  & 1.4E$-$5   & 2.09$^{+0.55}_{-0.53}$ &    \\
40706-01-01-000   & 51433 & 21 &  FB   &  2167  &  2.2  & 1.2E$-$3   & 2.45$^{+1.04}_{-1.07}$ &    \\ 
\tableline\tableline
\end{tabular}
\tablecomments{HB=Horizontal Branch, NB=Normal Branch, FB=Flaring Branch;
               Nu.= observation number}
\tablenotetext{a}{total live--time for HEXTE, in s}
\tablenotetext{b}{signal to noise ratio in the 75-220 keV energy range}
\tablenotetext{c}{probability that a better fit occurred due to a random chance}
\tablenotetext{d}{flux in the 75--220{\ts}keV energy range, in units of 
                  10$^{-10}$ ergs{\ts}cm$^{-2}${\ts}s$^{-1}$; apart from the 5 detections
                  of a hard X--ray tail, the photon index used for the non--thermal component
                  was our average value of 1; uncertainties are given at 90{\ts}{\%} confidence level}
\end{center}
\setlength{\extrarowheight}{0in}
}}
\end{table}

\clearpage

\begin{table}
\begin{center}
\caption{Observations of hard X--ray tail detections in Sco X--1 \label{tab2}}
\begin{tabular}{c c c c c c c}
\tableline
\tableline
                &\multicolumn{2}{c}{Bremsstrahlung}           &         &\multicolumn{2}{c}{Power--Law}                                                        &    \\
\cline{2-3}
\cline{5-6}
    OBSID       &  kT (keV)      &        Flux{\tablenotemark{a}} & &      $\Gamma$           &      Flux{\tablenotemark{b}}                                   & $\chi^2$ (dof)   \\
\tableline
20053-01-01-00  & 4.83$^{+0.04}_{-0.05}$ & 9.03$^{+0.39}_{-0.36}$ & & ~1.75$^{+0.22}_{-0.20}$ & 1.56$^{+0.10}_{-0.11}$          & 36.76 (23)  \\
20053-01-01-06  & 4.50$^{+0.05}_{-0.05}$ & 5.83$^{+0.30}_{-0.27}$ & & ~1.64$^{+0.29}_{-0.27}$ & 0.91$^{+0.11}_{-0.10}$          & 24.52 (23)  \\
20053-01-02-00  & 4.36$^{+0.03}_{-0.03}$ & 7.16$^{+0.26}_{-0.24}$ & & -0.17$^{+0.30}_{-0.33}$ & 1.22$^{+0.13}_{-0.14}$          & 29.15 (23)  \\
30036-01-01-000 & 4.28$^{+0.03}_{-0.04}$ & 9.63$^{+0.42}_{-0.41}$ & & -0.71$^{+0.63}_{-0.70}$ & 0.63$^{+0.11}_{-0.11}$          & 28.57 (22)  \\
30035-01-01-00  & 4.51$^{+0.08}_{-0.07}$ & 7.45$^{+0.60}_{-0.67}$ & & ~2.37$^{+0.33}_{-0.28}$ & 1.04$^{+0.08}_{-0.08}$          & 24.01 (22)  \\ \tableline
\end{tabular}
\tablecomments{uncertainties are given at 90{\ts}{\%} confidence level for the derived
parameters of the model applied}
\tablenotetext{a}{Flux in 20--50{\ts}keV range  in units of 10$^{-9}$ ergs{\ts}cm$^{-2}${\ts}s$^{-1}$}
\tablenotetext{b}{Flux in 20--200{\ts}keV interval  in units of
                  10$^{-9}$ ergs{\ts}cm$^{-2}${\ts}s$^{-1}$}
\end{center}
\setlength{\extrarowheight}{0in}
\end{table}

\clearpage

\figcaption[D'Amico et al., fig1.ps]
           {Spectrum resulting from a bremsstrahlung + power law
            fit to data subset 20053-01-02-00 (a), and 
            the result of a bremsstrahlung fit to subset
            30035-01-04-00 (b),
            showing the presence/absence of a hard X--ray tail in Sco X--1.
            Residuals are given in units of standard deviations (lower panels).  
            In (b) the upper limits are 2$\sigma$, including the 60--80{\ts}keV
            bin which experienced a $-${\ts}3$\sigma$ residual.
            The ${\chi}^2_{\nu}$ are 1.27 and 1.62 respectively.\label{fig1}}

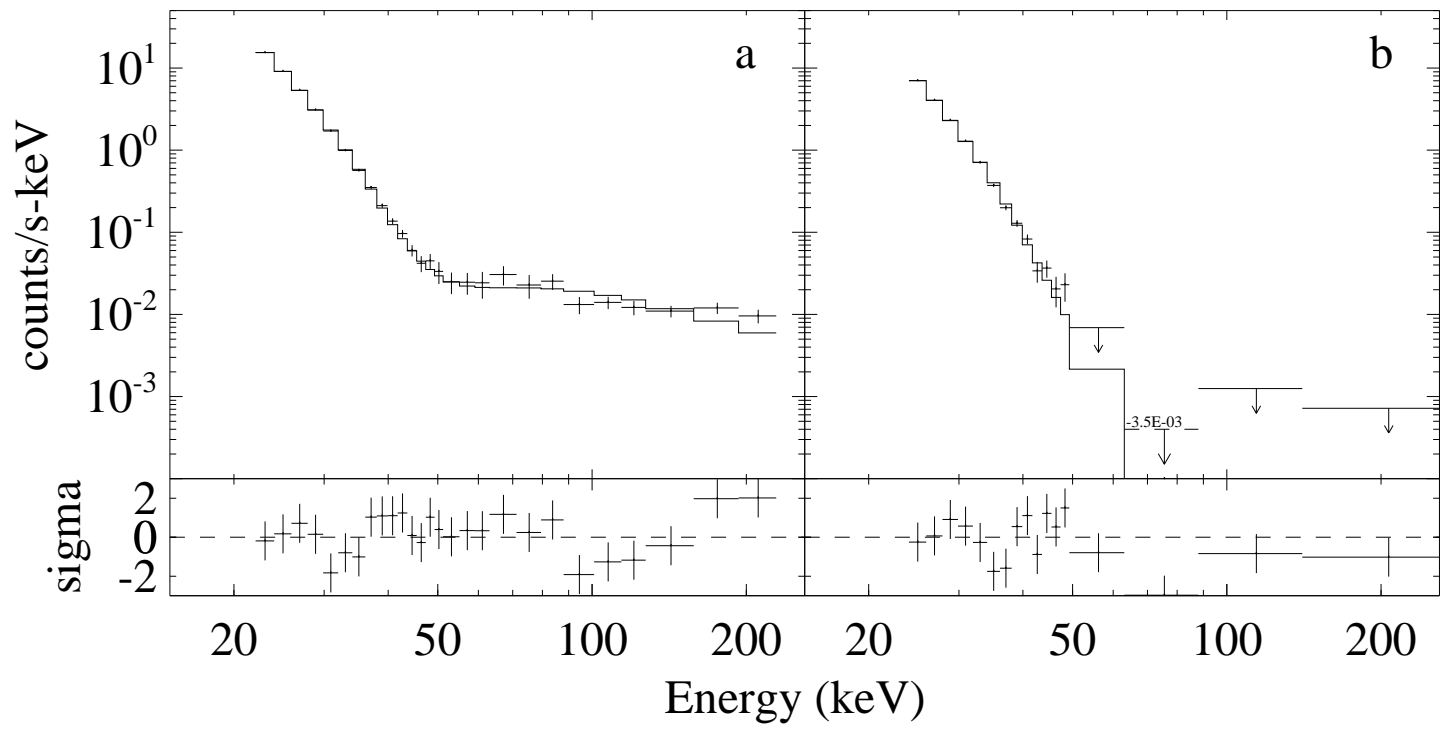
\end{document}